\documentstyle[12pt]{article}
\begin{document}

\thispagestyle{empty}
{}~ \hfill TECHNION-PH-96-4/REV\\
\vspace{1cm}

\begin{center}
{\large \bf  Theoretical Update on Two Non-Resonant 
Three-Body}\\[3mm]
{\large \bf Channels in Charmed Meson Decays}\\
\vspace{2.00cm}
{\bf  Da-Xin Zhang\footnote{e-mail:
zhangdx@phys1.technion.ac.il,
zhangdx@techunix.technion.ac.il}}\\
\vspace{0.30cm}
Department of Physics, Technion -- Israel Institute
 of Technology,\\
 Haifa 32000, Israel\\
\vspace{3.5cm}
{\bf Abstract}
\end{center}

\vspace{1.0cm}
\noindent 
Predictions of two channels in  the three-body
decays of the charmed mesons are made within the
heavy hadron chiral perturbation theory.
There still exists the problem that
the theoretical expectation is too small compared to
the experimental data.

\newpage

Nonleptonic weak decays of the charmed mesons
have been studied extentively in the past two decades.
Previous theoretical studies are focused on
the two-body cases\cite{2body}.
The experimental measurements have also
been achieved in some three-body channels\cite{ex-3body}.
The experimental results, however, are not well
understood quantatively due to two reasons.
On the theoretical side,
there exists no  method
in the literature which allows one to perform
the  calculation reliably.
On the experimental side, there are  so many open
resonant channels which contribute to the final
three-body states  that  the data are
difficult to be analysed.
\par

\vspace{0.4cm}
In the present work, we use the heavy hadron chiral
perturbation theory\cite{hhchpt,yan,dono}(HCHPT) to study the 
non-resonant three-kaon decays
of the charmed mesons.
In fact, application of  chiral
perturbation theory in this kind of study is not a new idea.
In the past, the $U(4)_L\otimes U(4)_R$ chiral symmetry 
has been  used in \cite{su4a,su4b}. 
Because this  symmetry is badly broken,
these predictions are totally not under control.
\par

\vspace{0.4cm}
HCHPT introduced in \cite{hhchpt}
can be described in the following.
The QCD lagrangian for the light quark
 ($u, d, s$) sector
possesses the $SU(3)_L\otimes SU(3)_R$ chiral symmetry.
The heavy quark ($b$ or $c$), which transforms as singlet under
the chiral symmtery, has the spin-flavor symmetry
in the limit that its mass is taken to be infinity\cite{hqs}.
 As the consequence,
the two lowest lying mesons containing one heavy
quark are degenerate in the heavy quark limit, and
can be expressed by the superfield
\begin{equation}
H_a=\frac{\sqrt{m_D}}{2}(1+\not v)(D_{a\mu}^{*}\gamma^{\mu}
-D_a^{}\gamma_5)~,
\label{superfield}
\end{equation}
where  use of the charmed mesons,
$c\bar{q}_a$ with $a=1, 2, 3$ corresponding to
$D^{0(*)}$, $D^{+(*)}$, $D^{+(*)}_s$, has been made 
as the example.
In (\ref{superfield}) we have suppressed the
explicit dependence of $H_a$ on its velocity $v$.
The strong interactions of the heavy mesons
with the pseudo-Goldstone bosons $\pi$, $K$ and $\eta$
at low energy can be constructed by taking
the derivative expansions on the pseudo-Goldstone
field $\Sigma=\xi^2=\exp(2iM/f)$, where
\begin{equation}
M=\left[ \begin{array}{ccc}
\frac{1}{\sqrt{2}}\displaystyle\pi^0+\frac{1}{\sqrt{6}}
\displaystyle\eta&\displaystyle\pi^+&K^+\\[3mm]
\displaystyle\pi^-&-\frac{1}{\sqrt{2}}\displaystyle\pi^0+\frac{1}
{\sqrt{6}}\displaystyle\eta&K^0\\[3mm]
K^-&\bar{K}^0&-\frac{2}{\sqrt{6}}\displaystyle\eta
\end{array}	\right]~,
\end{equation}
and $f$ is the decay constant of  pseudo-Goldstone bosons. 
In the derivative expansions, higher order terms
are suppressed by powers of $1/\Lambda_{CSB}$ with
the  chiral symmetry breaking scale
$\Lambda_{CSB}\sim 1.2{\rm GeV}$
from the naive dimensional analysis\cite{nda}.
As the superfield (\ref{superfield}) is used,
the requirements of the heavy quark symmetry 
is satisfied automatically.
Higher order terms which violate the heavy quark symmetry
are suppressed by powers of $1/M_Q$ and can be incorporated
into HCHPT\cite{1overm}.
\par

\vspace{0.4cm}
To the leading order in both the 
derivative and the $1/M_Q$ expansions,
the effective lagrangian in  HCHPT is
\begin{equation}
\begin{array}{rcl}
{\cal L}&=&-iTr \bar{H}_av_{\mu}\partial^{\mu} H_a+
\frac{1}{2}iTr \bar{H}_aH_bv_{\mu}(\xi^
+\partial^{\mu}\xi+
\xi\partial^{\mu}\xi^+)_{_{ba}}\\[3mm]
& &+\frac{1}{2}igTr \bar{H}_aH_b\gamma_{\mu}
\gamma_5 (\xi^+\partial^{\mu}\xi-
\xi\partial^{\mu}\xi^+)_{_{ba}}+\cdots~,
\end{array}
\label{leff}
\end{equation}
where the trace is taken over the spinor space.
The coupling $g$ in (\ref{leff}) is estimated to be of  order one
and can be extracted from the partial  width
of  the strong decays $D^*\rightarrow D\pi$.
Up to now it has 
only an upper bound $|g|\leq 0.63$\cite{g}. 
We will use $g=0.5\pm 0.1$
in the numerical estimations.
\par

\vspace{0.4cm}
In HCHPT, semiloptonic  decays of heavy-to-light
transitions are descibed by the effective weak current
\begin{equation}
\bar{q}_a\gamma_\mu (1-\gamma_5)c
=\displaystyle\frac{i\alpha}{2}Tr \gamma_\mu (1-\gamma_5)H_b
\xi^\dagger_{ba},
\label{curreff}
\end{equation}
where both sides transform as $(3_L,1_R)$ under
the $SU(3)_L\otimes SU(3)_R$ chiral symmetry, and
\begin{equation}
\alpha=f_D\sqrt{m_D}.
\end{equation}
The light quark current is described in the same way 
as in the usual chiral perturbation theory\cite{chpt}:
\begin{equation}
\bar{q}_i\gamma^\mu (1-\gamma_5)q_j
=\sum_{k}
\displaystyle\frac{if^2}{2}\partial^\mu\Sigma_{ik}\Sigma^\dagger_{kj}.
\label{lcurreff}
\end{equation}
\par

\vspace{0.4cm}
The effective hamiltonian for
Cabibbo-allowed three-$K$ decays of the charmed mesons is:
\begin{equation}
{\cal H}_{eff}=\displaystyle\frac{G_F}{\sqrt{2}}V_{cs}V_{ud}^*
[a_1\bar{u}\gamma^\mu(1-\gamma_5)d
\bar{s}\gamma_\mu(1-\gamma_5)c
+a_2\bar{s}\gamma^\mu(1-\gamma_5)d
\bar{u}\gamma_\mu(1-\gamma_5)c],
\label{heff}
\end{equation}
where $a_1=1.2$ and $a_2=-0.5$, which  are the most
favored values in the phenomenological analyses
in the two-body decays\cite{bsw},
will be used in numerically estimations.
In dealing with the nonleptonic decay ampitudes we use 
the factorization ansatz\cite{bsw}
under which the amplitudes for the three-body decays
depicted in Figure 1  for  $D^0\rightarrow K^+K^-\bar{K}^0$
and  $D^+\rightarrow K^+\bar{K}^0\bar{K}^0$ are:
\begin{equation}
\begin{array}{rcl}
<K^+(q_+)K^-(q_-)\bar{K}^0(q_0)|{\cal H}_{eff}|D^0>&=&
\displaystyle\frac{G_F}{\sqrt{2}}V_{cs}V_{ud}^*\\
&&\hspace{-6.0cm}[a_1
<K^+(q_+)\bar{K}^0(q_0)|\bar{u}\gamma^\mu(1-\gamma_5)d|0>
<K^-(q_-)|\bar{s}\gamma_\mu(1-\gamma_5)c|D^0>\\
&&\hspace{-6.0cm}+a_2
<\bar{K}^0(q_0)|\bar{s}\gamma^\mu(1-\gamma_5)d|0>
<K^+(q_+)K^-(q_-)|\bar{u}\gamma_\mu(1-\gamma_5)c|D^0>]
\\[5mm]
<K^+(q_+)\bar{K}^0(q_1)\bar{K}^0(q_2)|{\cal H}_{eff}|D^+>&=&
\displaystyle\frac{G_F}{\sqrt{2}}V_{cs}V_{ud}^*\sqrt{\displaystyle\frac{1}{2}}\\
&&\hspace{-6.0cm}[a_1
<K^+(q_+)\bar{K}^0(q_1)|\bar{u}\gamma^\mu(1-\gamma_5)d|0>
<\bar{K}^0(q_2)|\bar{s}\gamma_\mu(1-\gamma_5)c|D^+>\\
&&\hspace{-6.0cm}+a_2
<\bar{K}^0(q_1)|\bar{s}\gamma_\mu(1-\gamma_5)d|0>
<K^+(q_+)\bar{K}^0(q_2)|\bar{u}\gamma^\mu(1-\gamma_5)c|D^+>
\\
&&\hspace{-6.0cm}+~ q_1\Leftrightarrow q_2~ ].
\end{array}
\label{ampli}
\end{equation}
In Figure 1 we have discarded the W-exchange 
and the W-annihilation diagrams which are
expected to be highly suppressed.
\par

\vspace{0.4cm}
The applicability of the chiral perturbation theory
in $D\rightarrow 3K$ lies in the following reasons.  
In the final states, 
the maximum energy of each of the $K$-meson in the rest-frame
of the $D$ meson is
\begin{equation}
E_{max}\sim 0.73{\rm GeV},
\end{equation}
while the maximum value of the invariant mass  of any
two $K$-mesons is
\begin{equation}
(\sqrt{M_{ij}^2})_{max}=m_D-m_K\sim 1.3{\rm GeV},
\end{equation}
which is a little larger than the
estimation of  $\Lambda_{CSB}\sim 1.2$GeV 
from the naive dimensional analysis\cite{nda}.
However, $\Lambda_{CSB}$
can be  also  taken as $1.5{\rm GeV}$,
as has been analysed in the literature\cite{lambda}.
In this
case, the whole phase space of these decays are
in the region where HCHPT is applicable.
On the other hand,
even if $\Lambda_{CSB}\sim 1.2{\rm GeV}$ is taken, 
the  phase space where  HCHPT can be applied is still
dominant,
because it corresponds to the much large area
in the Dalitz plot.
Note that  discarding of  the annihilation diagram
is important to avoid the terms proportional to
the invariant mass of the three final particles.
\par

\vspace{0.4cm}
Note that the two channels depicted in Figure 1 are the only
three-body ones which can be analysed in HCHPT.
The  corresponding hadronic matrix elemnets in (\ref{ampli})
are estimated by calculating the Feynman diagrams in HCHPT
which  are depicted in Figure 2.
The results are
\begin{equation}
\begin{array}{rcl}
<K^+(q_+)\bar{K}^0(q_0)|\bar{u}\gamma^\mu(1-\gamma_5)d|0>
&=&(q_+-q_0)^\mu,\\
<\bar{K}^0(q_0)|\bar{s}\gamma^\mu(1-\gamma_5)d|0>&=&
iq_{0}^\mu f,\\
<K^-(q_-)|\bar{s}\gamma_\mu(1-\gamma_5)c|D^0>
&=&Y_1(q_-)_{\mu}+Y_2(q_-)_{\mu},\\
<K^+(q_+)K^-(q_-)|\bar{u}\gamma_\mu(1-\gamma_5)c|D^0>
&=&X_1(q_+,q_-)_{\mu}+X_2(q_+,q_-)_{\mu}\\
&+&X_3(q_+,q_-)_{\mu}+X_4(q_+,q_-)_{\mu},\\
[5mm]
<K^+(q_+)\bar{K}^0(q_1)|\bar{u}\gamma^\mu(1-\gamma_5)d|0>
&=&(q_+-q_1)^\mu,\\
<\bar{K}^0(q_1)|\bar{s}\gamma^\mu(1-\gamma_5)d|0>
&=&iq_{1}^\mu f,\\
<\bar{K}^0(q_2)|\bar{s}\gamma_\mu(1-\gamma_5)c|D^+>
&=&Y_1(q_2)_{\mu}+Y_2(q_2)_{\mu},\\
<K^+(q_+)\bar{K}^0(q_2)|\bar{u}\gamma_\mu(1-\gamma_5)c|D^+>
&=&X_1(q_+,q_2)_{\mu}+X_2(q_+,q_2)_{\mu}\\
&+&X_3(q_+,q_2)_{\mu}+X_4(q_+,q_2)_{\mu},
\end{array}
\label{feyndiag}
\end{equation}
where
\begin{equation}
\begin{array}{rcl}
X_1(q_+,q)_{\mu}&=&
i\displaystyle\frac{f_{D}P_{D\mu}}{f^2}
\displaystyle\frac{v\cdot(q-q_+)}{2v\cdot (q+q_+)},
\\[4mm]
X_2(q_+,q)_{\mu}&=&
-ig^2\displaystyle\frac{f_{D}P_{D\mu}}{f^2}
\displaystyle\frac{q\cdot q_+-v\cdot qv\cdot q_+}
{v\cdot (q+q_+)(v\cdot q+\Delta)},\\[4mm]
X_3(q_+,q)_{\mu}&=&
g\displaystyle\frac{-if_{D_s}}{f^2}
\displaystyle\frac{-m_Dq_\mu+v\cdot qP_{D\mu}}{v\cdot q+\Delta},
\\[4mm]
X_4(q_+,q)_{\mu}&=&\displaystyle\frac{if_{D}P_{D\mu}}{2f^2}
\end{array}
\end{equation}
coming from Figure 2(a)-(d), respectively, and
\begin{equation}
\begin{array}{rcl}
Y_1(q)_\mu&=&g\displaystyle\frac{f_{D_s}}{f}
\displaystyle\frac{-m_Dq_\mu+v\cdot qP_{D\mu}}{v\cdot q+\Delta},\\[4mm]
Y_2(q)_\mu&=&-\displaystyle\frac{f_{D_s}}{f}P_{D\mu}
\end{array}\end{equation}
from Figure 2(e)-(f).
We have denoted
\begin{equation}
\Delta=m_{D_s^*}-m_D.
\end{equation}
\par

\vspace{0.4cm}
In the numerically evaluations, we take
$f_D=f_{D_s}=0.2{\rm GeV}$ and
$f=f_K=0.161{\rm GeV}$.
The effective coupling
$g$ is taken to be $0.4$, $0.5$ or $0.6$(the  corresponding 
formfactor at the  maximum momentum transfer
in the semileptonic decays of $D\to K$ is
$|f_+(q_m^2)|=1.08$, $1.20$ or $1.38$,
while the experimental value is $1.30\pm 0.5$
if a monopole behavior of the $q^2$-dependence is used\cite{pdg}).
We give our results in Table 1, together with
the comparisions with  both the estimations 
from $U(4)_L\otimes U(4)_R$\cite{su4b} 
and the experimental data\cite{pdg}.
Note that  no numerical prediction 
has been made in \cite{su4a}
for the two channels we have studied.
As has been found in  
the  $U(4)_L\otimes U(4)_R$ studies\cite{su4b},
there are some three-body channels whose
measured branching ratios 
are larger than the theoretical expectations
by more than one  order.
In the two  channels  studied in the present work,
we are still suffered from the
same problem even if our calculations
are based on more reliable foundation.
This problem cannot be solved by going to
the higher order expansions in  HCHPT, otherwise
the expansions will not converge.
It  is also impossible to attribute
this  problem to the omissions of the W-annihilation
and the W-exchange diagrams because
they are suppressed compared to those in Figure 1.
\par

\vspace{0.4cm}
To bridge the gaps between the theoretical estimations
and the experimental measurements,
further studies at the future $\tau-Charm$
factory are essential, where strict subtractions
off the contributions from many resonant channels
should be carried out. 
In the meantime, the interference effects
between resonant and non-resonant channels 
are also needed to be studied  by both 
the theoriests and  the experimentists.
\par

\vspace{0.5cm}
The author would like to acknowledge G. Eilam for  suggestion of the present
work and helpful discussions.
This research  is supported in part by Grant 5421-3-96
from the Ministry of
Science and the Arts of Israel.

\newpage

\vskip 2.0cm
\centerline{\bf\large Table}
\vskip 1.0cm

\noindent
{\bf Table 1}
Comparisions of numerical results.
Our results using different values of $g$ are given
in the second column.
\\[10mm]
\begin{tabular}{c||ccc|c|c} 
\hline\hline
process&g=0.4&g=0.5&g=0.6&
\cite{su4b}&Exper.\cite{pdg}\\
\hline\hline
Br($D^0\rightarrow K^+K^-\bar{K}^0$)
&$1.3\times 10^{-4}$ &$1.8 \times 10^{-4}$ &$ 2.3\times 10^{-4}$
&$6.3\times 10^{-5}$
&
$\begin{array}{c}
(4.9\pm 0.9)\times 10^{-3}\\[-0.2cm]
(non-\phi)
\end{array}$
\\
\hline
Br($D^+\rightarrow K^+\bar{K}^0\bar{K}^0$)
&$6.4\times 10^{-4}$&$8.3\times 10^{-4}$&$1.1\times 10^{-3}$
&$-$&$(3.1\pm 0.7)\%$\\
\hline\hline
\end{tabular}

\vskip 4.0cm
\centerline{\bf\large Figures}

\vskip 1.0cm
\noindent {\bf Figure 1}
The Feynman diagrams for the  $D^0\rightarrow K^+K^-\bar{K}^0$
and  $D^+\rightarrow K^+\bar{K}^0\bar{K}^0$.

\vskip 1.0cm
\noindent {\bf Figure 2}
The Feynman diagrams used in HCHPT to calculate
the hadronic matrix elments between the heavy and the light
mesons.

\input epsf
\newpage
\begin{figure}[htb]
\centerline{\epsfxsize3.8in\epsfbox{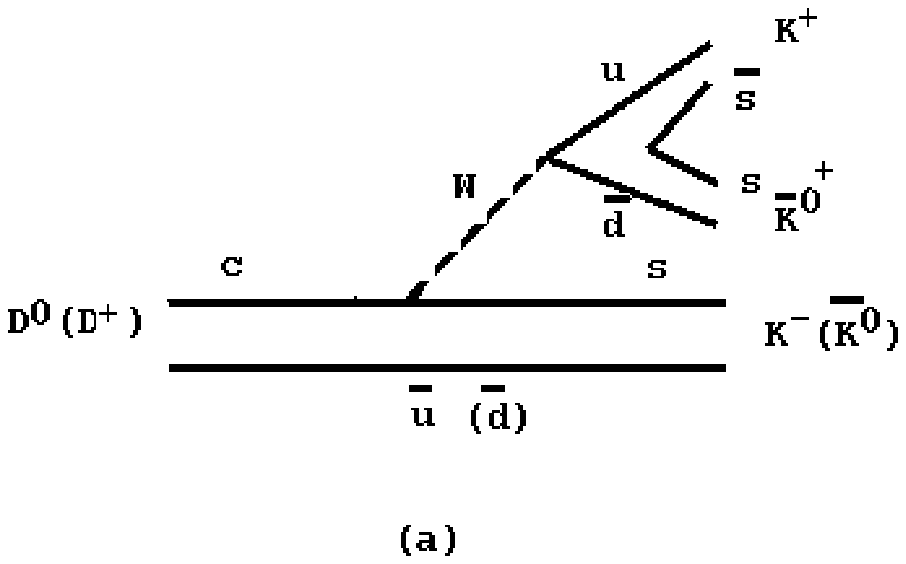}}
\centerline{\epsfxsize3.5in\epsfbox{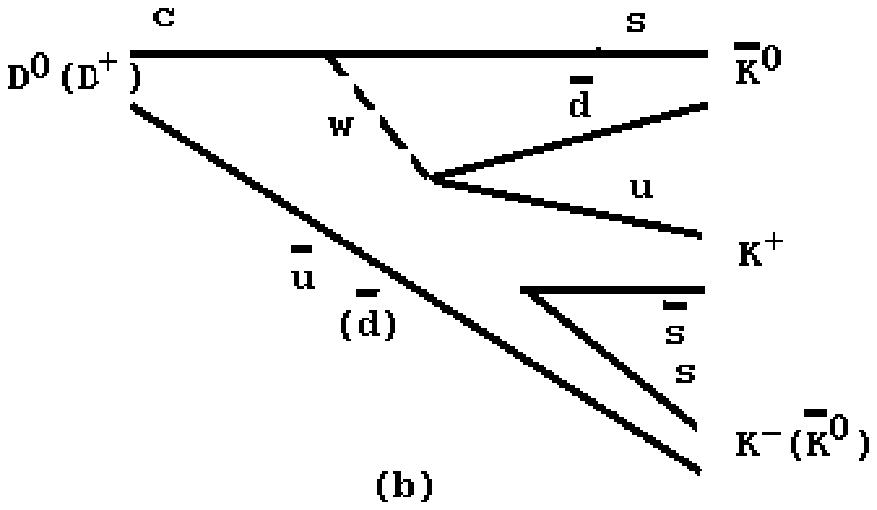}}
\caption{}
\end{figure}

\newpage
\begin{figure}[htb]
\centerline{\epsfxsize3.5in\epsfbox{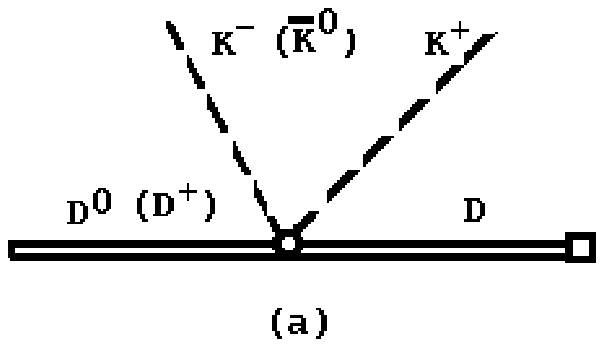}}
\centerline{\epsfxsize3.3in\epsfbox{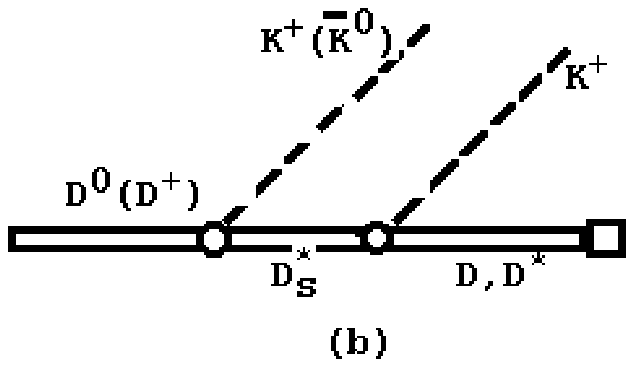}}
\centerline{\epsfxsize3.3in\epsfbox{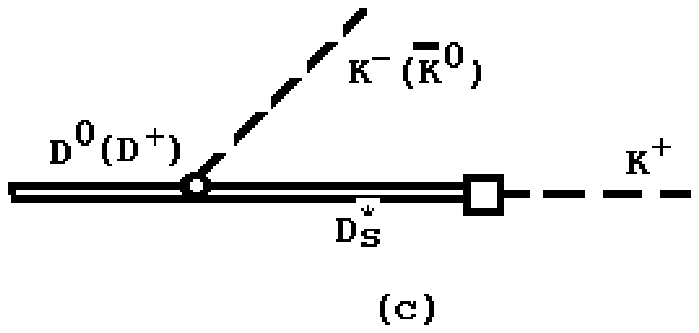}}
\end{figure}

\newpage
\begin{figure}[htb]
\centerline{\epsfxsize3.5in\epsfbox{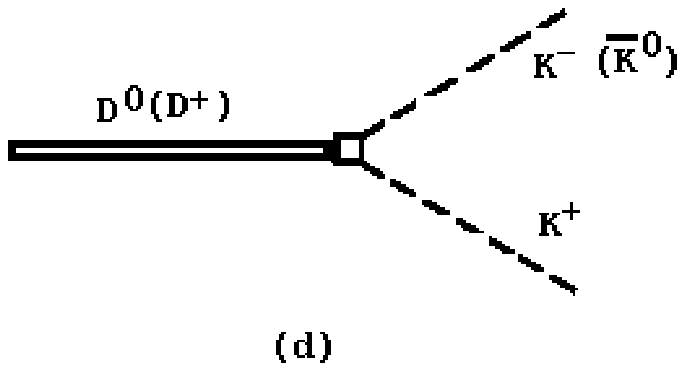}}
\centerline{\epsfxsize3.3in\epsfbox{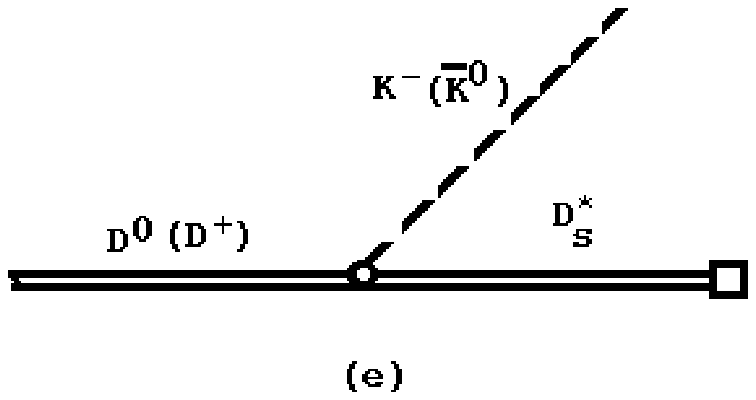}}
\centerline{\epsfxsize3.0in\epsfbox{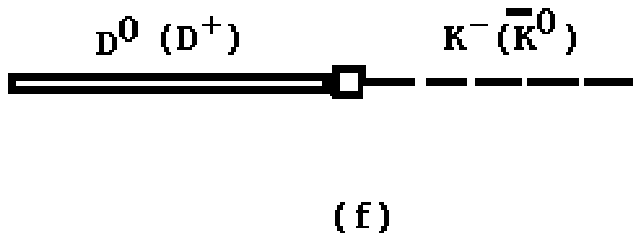}}
\caption{}
\end{figure}


\newpage


\begin{thebibliography}{99}

\bibitem{2body}
For a review, see M.  Wirbel, 
Prog. in Part. and Nucl. Phys.  21, 33 (1989);\\
A. J. Buras and M. K. Harlander, in {\it Heavy Flavors},
Eds. A.J. Buras, M. Lindner, World Scientific, 1992.

\bibitem{ex-3body}
R. Ammar {\it et al.} (CLEO Collaboration), Phys. Rev. D44, 3383 (1991);\\
P. L. Fradetti 
{\it et al.} (E687 Collaboration), Phys. Lett. B286, 195 (1992).

\bibitem{hhchpt}
 M. B. Wise, Phys. Rev. D45, 2188 (1992). For a review,
see M. B. Wise, Lectures given at the CCAST
Symposium on Particle
Physics at the Fermi Scale (1993), preprint CALT-68-1860.

\bibitem{yan}
T.-M. Yan, H.-Y. Cheng, G.-L. Lin, Y.C. Lin and H.-L. Yu,
Phys. Rev. D46, 1148 (1992).

\bibitem{dono}
G. Burdman and J. Donoghue, Phys. Lett. B280, 287 (1992);\\
 P. Cho, Phys. Lett. B285, 145 (1992).

\bibitem{su4a}
L.-L. Chau and H.-Y. Cheng, Phys. Rev. D41, 1510 (1990).

\bibitem{su4b}
F.J. Botella, S. Noguera and J. Portoles,
Phys.Lett.B360, 101 (1995).

\bibitem{hqs}
N. Isgur and M. Wise, Phys. Lett. B232, 113 (1989);
B237, 527 (1990);\\
E. Eichten and B.  Hill, Phys.Lett. B234, 511 (1990);\\
B. Grinstein, Nucl.Phys.B339, 253 (1990);\\
H. Georgi, Phys. Lett. B240, 447 (1990).

\bibitem{nda}
A. Manohar and H. Georgi, Nucl. Phys. B234, 189 (1984);\\
H. Georgi and L. Randall, Nucl. Phys. B276, 241 (1984);\\
H. Georgi, Phys. Lett. B298, 187 (1993).

\bibitem{1overm}
H.-Y. Cheng, C.-Y. Cheung, G.-L. Lin, Y.C. Lin,
T.-M. Yan and H.-L. Yu,
Phys.Rev.D49, 2490 (1994);\\
N.  Kitazawa and T. Kurimoto, Phys.Lett.B323, 65 (1994);\\
N. Di Bartolomeo, R. Gatto, F. Feruglio and G. Nardulli,
Phys.Lett.B347, 405 (1995).

\bibitem{g}
S. Barlag {\it et al.} (ACCMOR Collaboration),
Phys. Lett. B278, 480 (1992).

\bibitem{chpt} 
S. Weinberg, Physica 96A, 327 (1979);\\
J. Gasser and  H. Leutwyler, Ann. Phys. (NY) 158, 142 (1984).\\
See also the books
Howard Georgi, "Weak Interactions and Modern Particle Theory"
(Benjamin/Cummings, 1984), 
and John F. Donoghue, Eugene Golowich, Barry R. Holstein,
"Dynamics of the Standard Model" (Cambridge Univ. Press, 1992).

\bibitem{bsw}
M. Bauer, B. Stech and M. Wirbel, Z.Phys.C34, 103 (1987).

\bibitem{lambda}
J. Gasser and H. Leutwyler, Ann. of Phys. {158}, 142 (1984);
Nucl. Phys. B {250}, 465 (1985);\\
J. L. Goity, Phys. Rev. D {46}, 3929 (1992);\\
D. Du, C. Liu and D.-X. Zhang, Phys. Lett. B317, 179 (1993).

\bibitem{pdg}
Particle  Data  Group,  Phys. Rev. D50, 3$-I$ (1994).

\end{thebibliography}
\end{document}